\documentclass[%
 reprint,
 amsmath,amssymb,
 aps,
]{revtex4-1}

\usepackage{graphicx}
\usepackage{dcolumn}
\usepackage{bm}
\usepackage{hyperref}
\usepackage{siunitx}

\begin{document}

\preprint{APS/123-QED}

\title{Spectrally Broadband Electro-Optic Modulation with Nanoelectromechanical String Resonators}

\author{Nicolas Cazier}
\author{Pedram Sadeghi}
\author{Miao-Hsuan Chien}
\author{Mostafa Moonir Shawrav}
\author{Silvan Schmid}%
 \email{silvan.schmid@tuwien.ac.at}
\affiliation{%
 Institute of Sensor and Actuator Systems, TU Wien, 1040 Vienna, Austria.
}%

\date{\today}

\begin{abstract}

In this paper, we present an electro-optical modulator made of two  parallel nanoelectromechanical silicon nitride  string resonators. These strings are covered with electrically connected gold electrodes and actuated either by Lorentz or electrostatic forces. The in-plane string vibrations modulate the width of the gap between the strings. The gold electrodes on both sides of the gap act as a mobile mirror that modulate the laser light that is focused in the middle of this gap. These electro-optical modulators can achieve an optical modulation depth of almost 100$\%$ for a driving voltage lower than 1 mV at a frequency of 314 kHz. The frequency range is determined by the string resonance frequency, which can take values of the order of a few hundred kilohertz to several megahertz. The strings are driven in the strongly nonlinear regime, which allows a frequency tuning of several kilohertz without significant effect on the optical modulation depth.

\end{abstract}


\maketitle

\section{Introduction}

In recent years, optomechanical systems have generated a lot of interest \cite{Favero2009, Gomis-Bresco2014, Aspelmeyer2014}, in part because of their many applications as sensors for precision measurements \cite{Bagci2014, Miao2019}, but also because of their usefulness as reconfigurable metamaterials \cite{Ou2013, Ou2016, Dong2016, Valente2014, Zheludev2016, Karvounis2019} and as plasmomechanical resonators \cite{Thijssen2013, Thijssen2014, Thijssen2015, Schmid2014, Naumenko2016, Herrmann2016, Roxworthy2016}. Among nanomechanical resonators, silicon nitride (SiN) strings and membranes stand out because of their very high quality factors, which make them very useful for experimenting on cavity optomechanics \cite{Wilson2009, Yamada2013, Schmid2014a}, or for designing sensors \cite{Roxworthy2016, Schmid2014, Gavartin2012, Larsen2013} and optomechanical systems \cite{Thijssen2015, Roxworthy2016, Schmid2014, Thijssen2013}.

In particular, reconfigurable metamaterials made of arrays of silicon nitride string resonators have been demonstrated as effective electro-optic modulators, where the strings were actuated using either electrostatic forces \cite{Ou2013}, Lorentz forces \cite{Valente2014}, or electrostriction \cite{Karvounis2019}. However, these metamaterial-based electro-optical modulators suffer from having either low optical modulation depths \cite{Valente2014} or requiring high driving voltages \cite{Ou2013, Karvounis2019}.

In this paper, we present a different kind of electro-optical modulator, made of two gold covered and electrically connected silicon nitride string resonators, separated by a small gap in their center, similar to the structures used by Thijssen et al. in Ref. \cite{Thijssen2013}, but actuated electromagnetically using Lorentz forces to change the width of the gap between the strings, by driving one of the strings at the resonance frequency of its in-plane fundamental mode. When a laser is focused in the middle of the gap, we can use this to modulate the reflection of this laser on the gold electrode covering the string. This gives us an easy way to fabricate MEMS electro-optical modulators with an optical modulation depth that can reach almost 100$\%$ for a driving voltage below 1 mV. We also tested similar structures that were actuated using electrostatic forces (generated by comb-drive actuators between the strings) instead of Lorentz forces, and achieved an optical modulation depth of 82{\%} for a driving voltage of 250 mV for the electro-optical modulators based on these structures.

\section{Methods}

\begin{figure}
\begin{centering}
\includegraphics[width=8.25cm]{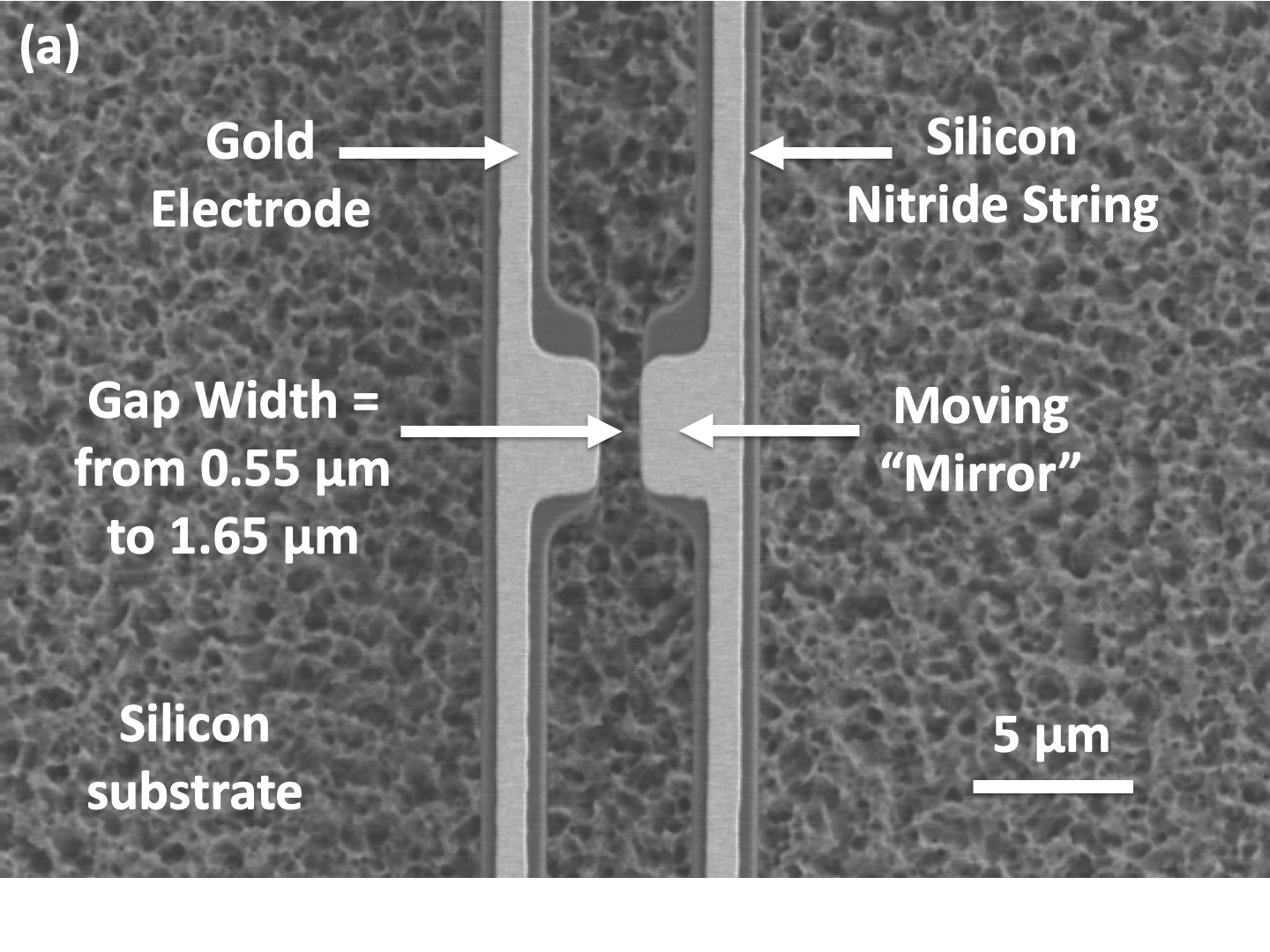} \includegraphics[width=8.2cm]{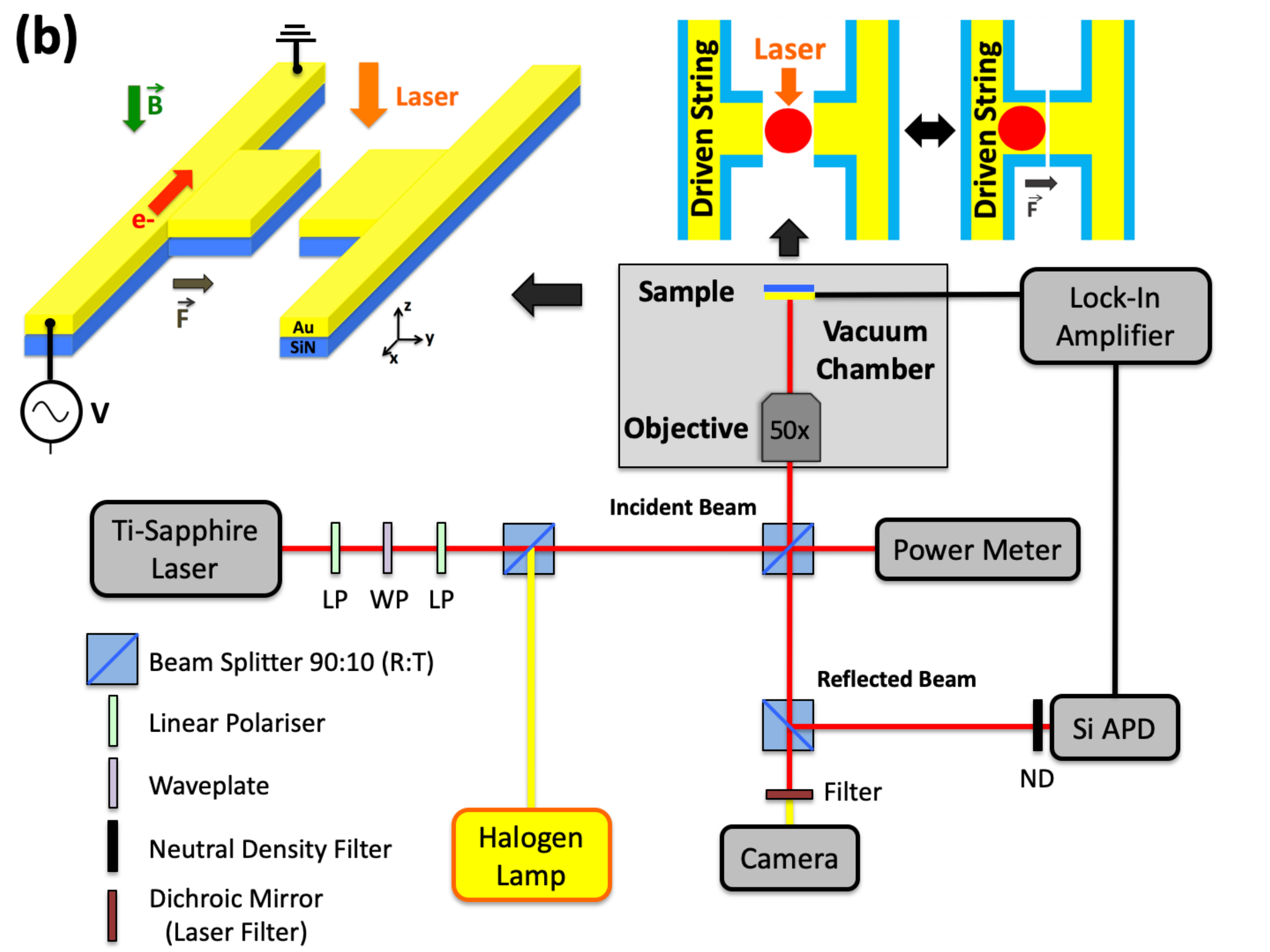}
\caption{\label{fig:figure1} (a) SEM Image of the opto-mechanical resonator, made of two SiN strings covered with electrically-connected gold electrodes. The width of the gap at the center of the strings is between 0.55 ${\mu}$m and 1.65 ${\mu}$m depending on the sample measured. (b) The experimental setup used for our optical measurements. The strings are placed in a vacuum chamber with a static magnetic field  (200 mT), and an oscillating current is sent through one of the strings. This will create Lorentz forces that will excite the in-plane mode of the string, changing the width of the gap between the strings. A tunable Ti-Sapphire laser, focused on the center of the gap between the strings, is then used to detect the string vibration through the modulation of the laser reflection on the moving mirror. A waveplate and two linear polarizers are used to adjust the laser power and its polarisation. The sample is kept in a vacuum chamber with the neodymium magnets used to create the static magnetic field. A silicon avalanche photodetector (Si APD) is used to detect the reflected optical signal. Both the APD and the sample are connected to a lock-in amplifier used to send the electrical current through the SiN string and to visualize the resulting changes in the reflected laser optical power measured by the APD.}
\end{centering}
\end{figure}

The wafer used to fabricate our optomechanical resonators was a 380 ${\mu}$m thick silicon wafer, covered on both sides by a 300 nm thick layer of low stress (200 MPa) LPCVD Si$_3$N$_4$. The fabrication process was then done in three steps. First we deposited the gold electrodes using UV lithography, followed by thermal evaporation of a 100 nm thick gold layer and then lift-off. In the second step, the silicon nitride strings were created by UV lithography, followed by Reactive Ion Etching (RIE) of the silicon layer. The last step was to use dry etching with XeFl$_2$ to etch the silicon substrate over a depth of 6 ${\mu}$m in order to release the Si$_3$N$_4$ strings. At the end of the fabrication process, the width of the gap between the strings was between 0.55 ${\mu}$m and 1.65 ${\mu}$m depending on the sample measured (see Fig.~\ref{fig:figure1}a). During all our measurements, the laser was initially focused on the middle of this gap between the two strings.

The experimental setup used can be seen on Fig.~\ref{fig:figure1}b. The laser used is a Titanium-Sapphire laser (SolsTiS from M Squared), whose wavelength can be tuned from 730 nm to 1000 nm. The wavelength chosen for our measurements was 730 nm. The sample is placed in a vacuum chamber between two neodymium magnets that create a static magnetic field of about B = 200 mT at the center of the sample. The waveplate and linear polarizers are used to adjust the laser optical intensity as well as its polarization, which we chose to be parallel to the strings during our measurements. The laser optical power when it reaches the sample is 38 $\mu$W. A lock-in amplifier (UHFLI from Zurich Instruments) is used to send an oscillating current through one of the strings and detect the resulting variation in the reflected laser optical power using a silicon avalanche photodetector (APD410A/M from ThorLabs). A halogen lamp (HL-2000-FHSA from Ocean Optics), used as a white light source, and a camera (EO-0413C from Edmund Optics) are used to obtain an image of the sample so we can adjust the laser position and focus on it. This white light source is then turned off during the actual measurements. A dichroic mirror (DMSP650T from ThorLabs) is placed before the camera to cut most of the laser power while letting the white light in, since the laser light would otherwise saturate the camera, and another optical neutral density filter (Thorlabs NE510B-B) is placed before the photodetector in order to keep the reflected laser power below the photodetector saturation level of 1.5 $\mu$W. 

\section{Results}

\begin{figure}
\begin{centering}
\includegraphics[width=8.25cm]{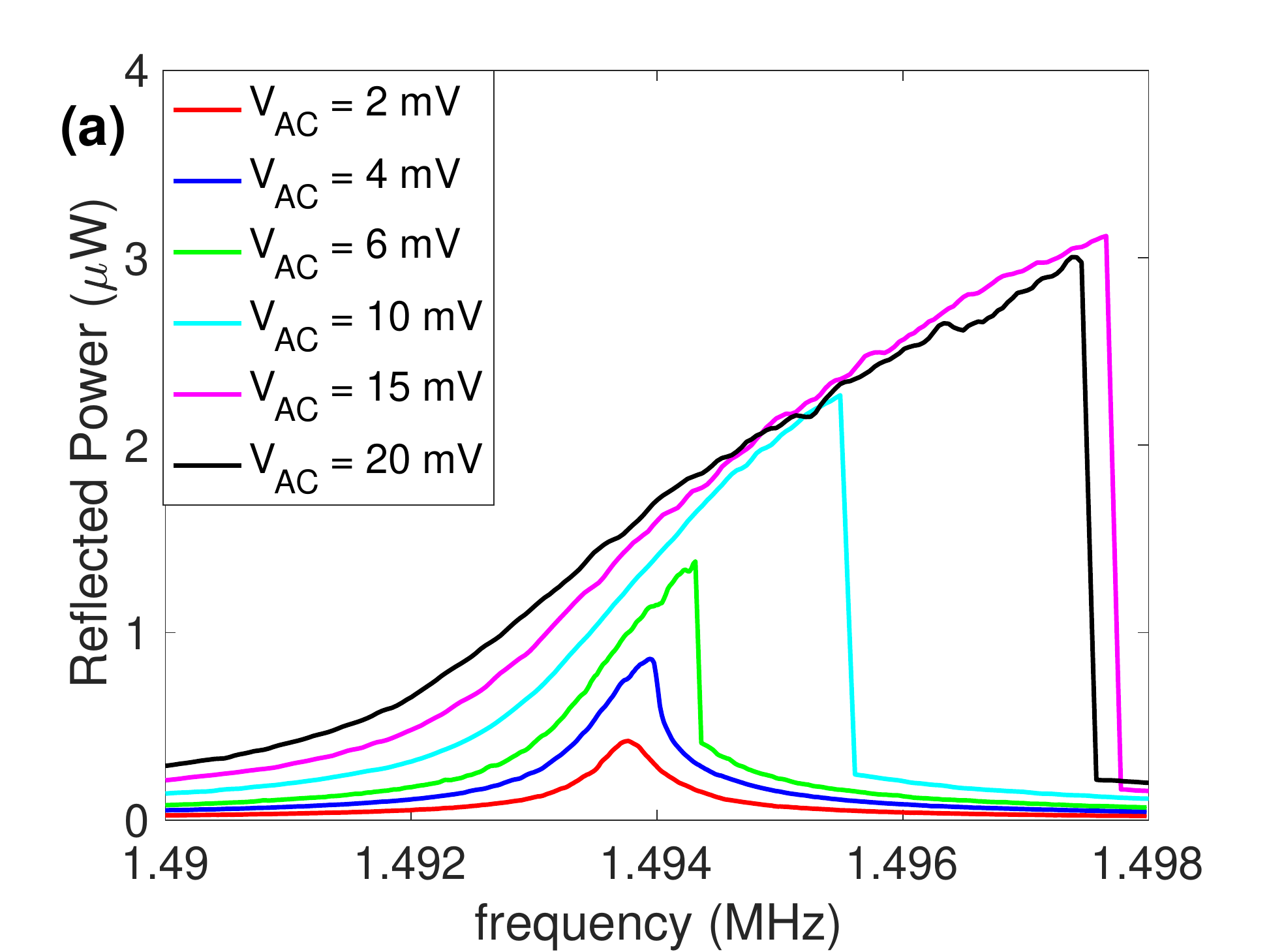} \includegraphics[width=8.25cm]{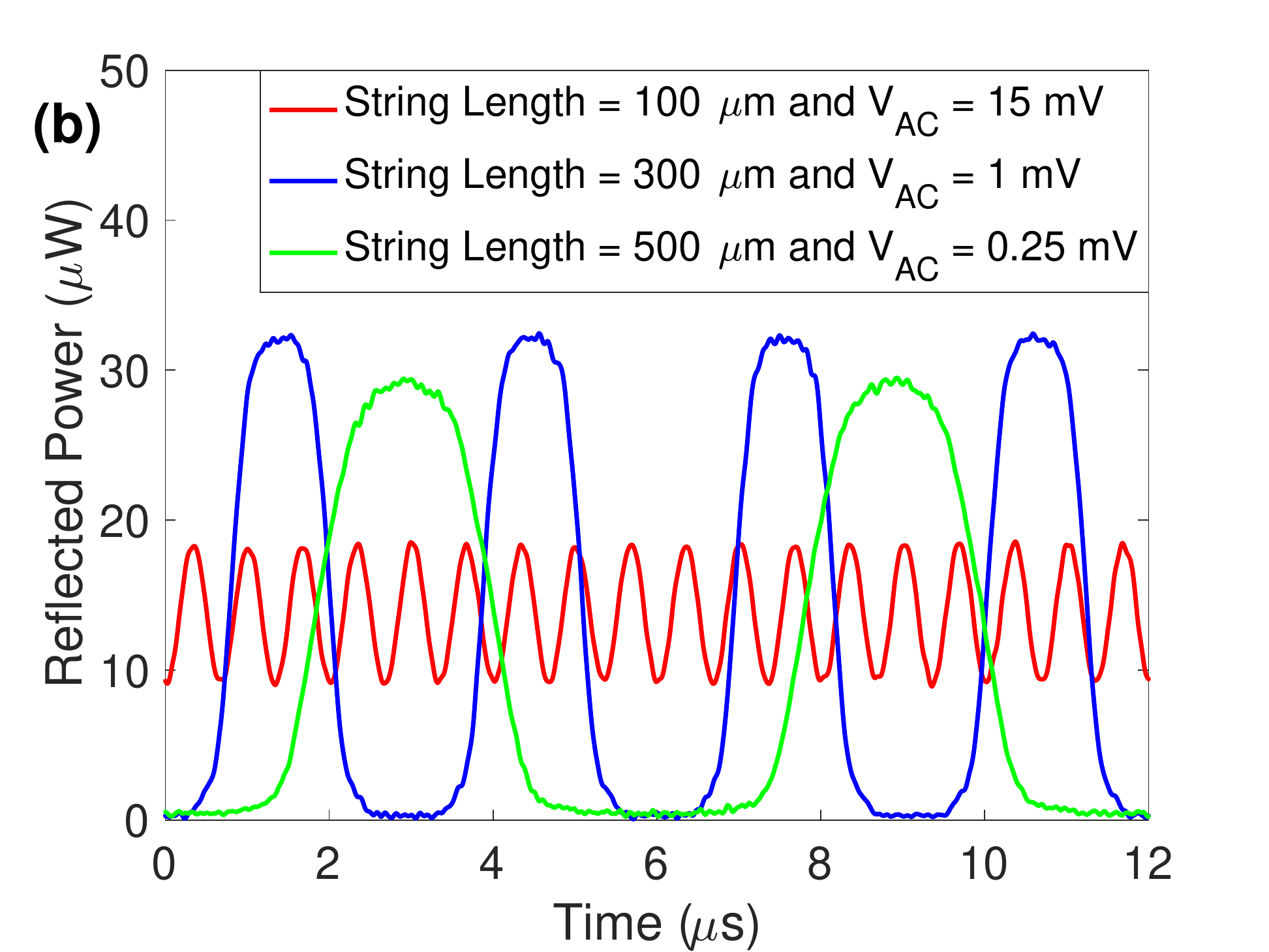}
\caption{\label{fig:figure2} (a) The mechanical resonance peak of the fundamental in-plane mode of one of the 100 $\mu$m long strings, as measured from the change in the reflected laser power when an alternating driving voltage is applied to the string. The resonance peak becomes strongly nonlinear for higher driving currents, which is typical of mechanical string resonators. However, there seems to be a maximal driving voltage (15 mV here) above which the shape and amplitude of the resonance peak no longer change significantly, maybe corresponding to the voltage where the gap between the strings is completely closed.  (b) The electro-optical modulation of the reflected laser power as a function of time, when driven at a frequency close to the resonance peak maximum, for the three samples we measured. These samples have different string lengths and therefore different resonance frequencies and maximal driving voltages. For the 100 $\mu$m string, we have a quasi-sinusoidal signal and an optical modulation depth of 31 ${\%}$. For the 300 $\mu$m and 500 $\mu$m strings, we are able to achieve an optical modulation depth of close to 100 ${\%}$ with a driving voltage below 1 mV, but the modulated signal is no longer quasi-sinusoidal.}
\end{centering}
\end{figure} 

We measured three different samples using Lorentz forces actuation, with different lengths for the string resonators. The first sample had strings with a length L = 100 $\mu$m and a gap width of 0.55 $\mu$m. Its fundamental resonance frequency for the in-plane mode is 1.494 MHz, and the resonance has a quality factor of 3000. The string resistance $R$ = 140 ${\Omega}$, which gives us, for a driving voltage $V_{AC}$ = 15 mV, a maximal Lorentz force $F_L = L \times B \times V_{AC}/R$ = 2.14 nN. The displacement of the string, as measured from the change in the reflected laser power, is then large enough to present a very strong mechanical nonlinearity, as can be seen on Fig.~\ref{fig:figure2}a, which shows a distortion of the resonance peak of the string typical of a nonlinear Duffing resonator \cite{Lifshitz2010}. However, unlike the theoretical nonlinear Duffing resonator, there seems to be a maximal driving voltage beyond which the resonance peak shape and amplitude no longer change. This maximal voltage is 15 mV here, but we observed a similar behavior for lower voltages in our other samples. We think this may correspond to the driving voltage where the gap between the strings close completely, which would prevent the string displacement from increasing even further. The width of the nonlinear resonance peak at higher driving voltages also means we can adjust the modulation frequency of our electro-optical resonator by a few kilohertz without losing much of the reflected signal amplitude.  

The red curve on Fig.~\ref{fig:figure2}b represents the electro-optical modulation of the reflected laser power as a function of time, measured when driving the string with a voltage $V_{AC}$ = 15 mV and a resonance frequency of 1.497 MHz, close to the peak maximum. The modulated signal here is quasi-sinusoidal, and has an optical modulation depth of 35.4 ${\%}$, which is the maximum we could obtain for this sample. However, we can obtain much higher optical modulation depths at lower driving voltages if we use longer strings because of their lower effective spring constant. For example, the second sample we measured had strings with a length of 300 $\mu$m, an electrical resistance of 250 ${\Omega}$ and a gap width of 1.65 $\mu$m. The fundamental resonance frequency for the in-plane mode is then 314 kHz, with a quality factor of 6800. When driving this string at its fundamental resonance frequency with a voltage $V_{AC}$ = 1 mV (corresponding to a Lorentz force F = 240 pN), we can then obtain an optical modulation depth of 99.6 ${\%}$ for the reflected laser power (see the blue curve on Fig.~\ref{fig:figure2}b), although the modulated signal is no longer quasi-sinusoidal here. We had similar results for our third measured sample, which had strings with a length of 500 $\mu$m, an electrical resistance of 360 ${\Omega}$ and a gap width of 1.6 $\mu$m. This time the fundamental resonance frequency for the in-plane mode is 166 kHz, with a quality factor of 7900, and we were able to obtain an optical modulation depth of 98.9 ${\%}$ (see the green curve on Fig.~\ref{fig:figure2}b) for the reflected laser power, with a driving voltage of only 0.25 mV (corresponding to a Lorentz force of 70 pN).

The maximal reflected laser power we measured (for the sample with the 300 $\mu$m long strings) is 32.1 $\mu$W for an incident laser power of 38 $\mu$W, which gives us a maximal reflection coefficient of 84.5 ${\%}$, close to the theoretical reflectivity of gold of 90 ${\%}$ for a wavelength of 730 nm \cite{Loebich1972}, so our electro-optical modulators seem to present very little losses. We can also note that the power consumption for these electro-optical modulators is very low : $P = \frac{1}{2}V_{AC}^2/R = 0.8$ $\mu$W for $V_{AC} = 15$ mV in the case of the 100 $\mu$m long strings, $P = 2$ nW for $V_{AC} = 1$ mV in the case of the 300 $\mu$m long strings and $P = 87$ pW for $V_{AC} = 0.25$ mV in the case of the 500 $\mu$m long strings.

If we compare our electro-optical modulators to reconfigurable metamaterial-based modulators, as e.g. \cite{Valente2014}, which are also actuated using Lorentz forces, we can note that we were able to improve the optical modulation depth from a few $\%$ up to 35 ${\%}$ to 100 ${\%}$ for a driving voltage of 15 mV to 0.25 mV. Also, unlike with metamaterial-based electro-optical modulators, whose transmission and reflection coefficients strongly depend on the laser wavelength, our electro-optical modulators will in theory have a reflection coefficient and an optical modulation depth that are quasi-independent of the laser wavelength used, at least for wavelengths above 650 nm where the reflectivity of gold is always above 90 $\%$ \cite{Loebich1972}.

\begin{figure}
\begin{centering}
\includegraphics[width=8.25cm]{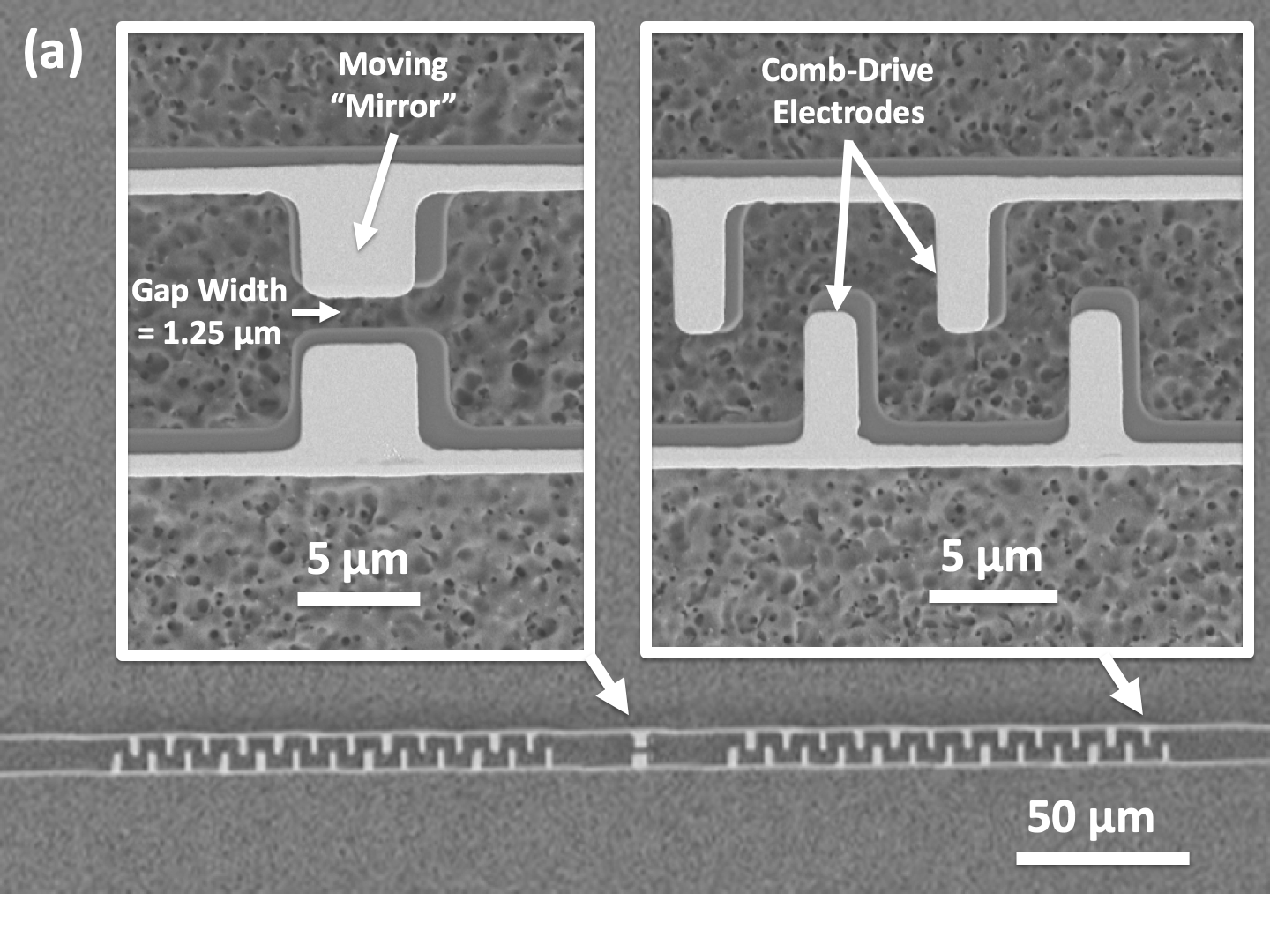} \includegraphics[width=8.25cm]{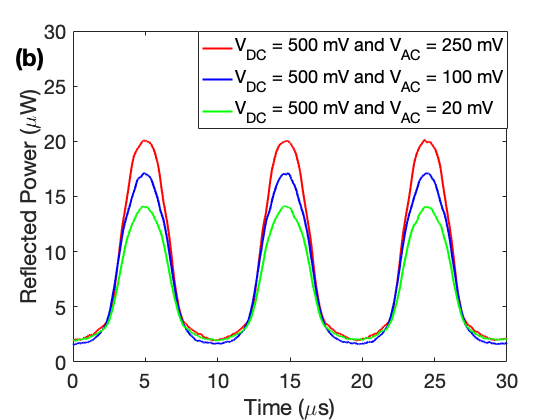}
\caption{\label{fig:figure3} (a) SEM image of the opto-mechanical resonator with with comb-drives actuators. The width of the gap at the center of the strings is 1.25 ${\mu}$m. (b) The electro-optical modulation of the reflected laser power as a function of time, when this resonator is driven at a frequency close to its resonance peak maximum, for different driving voltages. The optical modulation depth here is of 82.3 ${\%}$ for $V_{AC}$ = 250 mV.}
\end{centering}
\end{figure} 

To remove the bulky pair of magnets which is used to create the magnetic field for Lorentz force actuation, we also fabricated a sample with strings that could be driven with electrostatic forces using comb-drive actuators placed between the two strings (see Fig.~\ref{fig:figure3}a for the SEM picture), since this kind of electro-optical modulators would be easier to integrate. This sample has strings of length 600 $\mu$m with a gap width of 1.25 $\mu$m. The fundamental resonance frequency for the in-plane mode is 102 kHz and its quality factor is 9800. 

The electrostatic force created by the Comb-Drive actuators is theoretically equal to $F_{EL} = N\epsilon_0h_{Au}/D_{El}\times V^2$, with $N = 24$ the number of electrodes, $D_{El} = 3$ $\mu$m the distance between the electrodes, $h_{Au} = 100$ nm the thickness of the gold layer and $\epsilon_0 = 8.85$ $pF/m$ the vacuum permittivity \cite{Brand2015}. Since the electrostatic force is proportional to $V^2$, we cannot just apply directly a voltage $V = V_{AC}cos(\omega t)$ to our strings, as the force would be proportional to $V^2 = V_{AC}^2cos(\omega t)^2 = \frac{1}{2}V_{AC}^2(1+cos(2\omega t))$ and oscillate at twice the resonance frequency. By applying a DC voltage as well as a AC voltage like this : $V = V_{DC}+V_{AC}cos(\omega t)$, we can have $V^2 = V_{DC}^2+\frac{1}{2}V_{AC}^2+2V_{DC}V_{AC}cos(\omega t)+\frac{1}{2} V_{AC}^2cos(2\omega t)$, where the component of the electrostatic force in $cos(\omega t)$, the only one that can excite the fundamental in-plane mode, will be proportional to $2V_{DC}V_{AC}$. For $V_{DC} = 500$ mV and $V_{AC} = 250$ mV, this gives us an electrostatic force equal to $F_{EL} = N\epsilon_0h_{Au}/D_{El} \times 2V_{DC}V_{AC} = 1.77$ pN. This value is much lower than for the Lorentz forces, but unlike the Lorentz forces that are uniformly applied to the whole length of the string, the electrostatic forces are only applied to the middle part of the string and are therefore going to be much more efficient.

This sample was driven with a DC voltage $V_{DC} = 500$ mV and a AC voltage between 20 mV and 250 mV, this time applied to both ends of the electrode on one string, with the electrode on the other string being grounded, so no current was able to flow through the string's electrodes. Similarly to the previous samples actuated with Lorentz forces, we could easily detect an electro-optical modulation of the reflected laser power, with an optical modulation depth of 82.3 ${\%}$ for $V_{AC}$ = 250 mV (see Fig.~\ref{fig:figure2}b). However, like for our previous samples, there seems to be a specific driving voltage beyond which the maximal displacement of the string seems to saturate, or at least no longer increases linearly with the AC voltage used to drive the string, unlike what we would expect : When we increase $V_{AC}$ from 20 mV to 250 mV, the maximal amplitude of the reflected laser power only increases from 14 $\mu$W to 20 $\mu$W. This gives us a maximal reflection coefficient of 52.6 $\%$ for 22 $\mu$W, which is still below the maximal reflection coefficient of 84.5 $\%$ we measured for the samples using Lorentz forces actuation. Still, this is quite a good result, and shows that electrostatic forces would be a viable alternative to Lorentz forces for actuating these electro-optical modulators. Indeed, compared to similar reconfigurable metamaterial-based electro-optical modulators that are also actuated using electrostatic forces \cite{Ou2013}, our design constitutes a significant improvement in reversible modulation of the reflectance.

\section{Conclusion}

To summarize our results, we showed in this paper that electromagnetically-actuated, gold covered silicon nitride string resonators could be efficiently used to make MEMS electro-optical modulators, with an optical modulation depth easily reaching almost 100 $\%$ for a driving voltage below 1 mV and a power consumption below 2 nW. The modulation frequency is determined by the length of the string resonator, but can go from 100 kHz up to 1.5 MHz. However, the shorter string resonators needed for the higher modulation frequencies also have lower quality factors and therefore lower optical modulation depths, and will need higher driving voltages as well. These electro-optical modulators can be easily fabricated with standard microfabrication techniques, which would make them easy to integrate on-chip into a wider micro-opto-electro-mechanical system.

\begin{acknowledgments}

We would like to thank our technician Sophia Ewert for her cleanroom support. This work has also received funding from the European Research Council under the European Unions Horizon 2020 research and innovation program (Grant Agreement-716087-PLASMECS). 

\end{acknowledgments}



\begin{thebibliography}{10}

\bibitem{Favero2009}
Ivan Favero and Khaled Karrai.
\newblock {Optomechanics of Deformable Optical Cavities}.
\newblock {\em Nature Photonics}, 3(4):201--205, 2009.

\bibitem{Gomis-Bresco2014}
Jordi Gomis-Bresco, Daniel Navarro-Urrios, Mourad Oudich, Said El-Jallal,
  Amadeu Griol, Daniel Puerto, Emiglio Chavez, Yan Pennec, Bahram
  Djafari-Rouhani, Franscesca Alzina, Alejandro Mart{\'{i}}nez, and Clivia.
  M.~Sotomayor Torres.
\newblock {A 1D Optomechanical Crystal with a Complete Phononic Band Gap}.
\newblock {\em Nature Communications}, (5):4452, July 2014.

\bibitem{Aspelmeyer2014}
Markus Aspelmeyer, Tobias~J Kippenberg, and Florian Marquardt.
\newblock {Cavity Optomechanics}.
\newblock {\em Reviews of Modern Physics}, 86(4):1391--1452, 2014.

\bibitem{Bagci2014}
Tolga Bagci, Anne Simonsen, Silvan Schmid, Luis~Guillermo Villanueva, Emil
  Zeuthen, J.~Appel, Jacob~M. Taylor, Anders Sorensen, Koji Usami,
  A.~Schliesser, and Eugene~S. Polzik.
\newblock {Optical Detection of Radio Waves Through a Nanomechanical
  Transducer}.
\newblock {\em Nature}, 507(7490):81--85, March 2014.

\bibitem{Miao2019}
Markus Piller, Niklas Luhmann, Miao-Hsuan Chien, and Silvan Schmid.
\newblock {Nanoelectromechanical Infrared Detector}.
\newblock {\em Proceedings of SPIE}, 11088, 2019.

\bibitem{Ou2013}
Jun-Yu Ou, Eric Plum, Jianfa Zhang, and Nikolay~I. Zheludev.
\newblock {An electromechanically reconfigurable plasmonic metamaterial
  operating in the near-infrared}.
\newblock {\em Nature Technology}, 8(4):252--255, April 2013.

\bibitem{Ou2016}
Jun-Yu Ou, Eric Plum, Jianfa Zhang, and Nikolay~I. Zheludev.
\newblock {Giant Nonlinearity of an Optically Reconfigurable Plasmonic
  Metamaterial}.
\newblock {\em Advanced Materials}, 28:729--733, 2016.

\bibitem{Dong2016}
Biqin Dong, Xiangfan Chen, Fan Zhou, Chen Wang, Hao~F. Zhang, and Cheng Sun.
\newblock {Gigahertz All-Optical Modulation Using Reconfigurable Nanophotonic
  Metamolecules}.
\newblock {\em Nano Letters}, 2016.

\bibitem{Valente2014}
Joao Valente, Jun-Yu Ou, Eric Plum, Ian~J. Youngs, and Nikolay~I. Zheludev.
\newblock {A Magneto-Electro-Optical Effect in a Plasmonic Nanowire Material}.
\newblock {\em Nature Communications}, 6(7021), 2015.

\bibitem{Zheludev2016}
Nikolay~I. Zheludev and Eric Plum.
\newblock {Reconfigurable Nanomechanical Photonic Metamaterials}.
\newblock {\em Nature Nanotechnology}, 11(1):16--22, January 2016.

\bibitem{Karvounis2019}
Artemios Karvounis, Behrad Gholipour, Kevin~F. MacDonald, and Nikolay~I.
  Zheludev.
\newblock {Giant Electro-Optical Effect through Electrostriction in a
  Nanomechanical Metamaterial}.
\newblock {\em Advanced Materials}, 31:1804801, 2019.

\bibitem{Thijssen2013}
Rutger Thijssen, Ewold Verhagen, Tobias~J. Kippenberg, and Albert Polman.
\newblock {Plasmon Nanomechanical Coupling for Nanoscale Transduction}.
\newblock {\em Nano Letters}, 13(7):3293--3297, July 2013.

\bibitem{Thijssen2014}
Rutger Thijssen, Tobias~J. Kippenberg, Albert Polman, and Ewold Verhagen.
\newblock {Parallel Transduction of Nanomechanical Motion Using Plasmonic
  Resonators}.
\newblock {\em ACS Photonics}, 1(11):1181--1188, November 2014.

\bibitem{Thijssen2015}
Rutger Thijssen, Tobias~J. Kippenberg, Albert Polman, and Ewold Verhagen.
\newblock {Plasmomechanical Resonators Based on Dimer Nanoantennas}.
\newblock {\em Nano Letters}, 15(6):3971--3976, June 2015.

\bibitem{Schmid2014}
Silvan Schmid, Kaiyu Wu, Peter~Emil Larsen, Tomas Rindzevicius, and Anja
  Boisen.
\newblock {Low-Power Photothermal Probing of Single Plasmonic Nanostructures
  with Nanomechanical String Resonators}.
\newblock {\em Nano Letters}, 14(5):2318--2321, May 2014.

\bibitem{Naumenko2016}
Denys Naumenko, Valeria Toffoli, Silvio Greco, Simone {Dal Zilio}, Alpan Bek,
  and Marco Lazzarino.
\newblock {A Micromechanical Switchable Hotspot for SERS Applications}.
\newblock {\em Applied Physics Letters}, 109(13):131108, September 2016.

\bibitem{Herrmann2016}
Lars~O Herrmann, Antonis Olziersky, Cynthia Gruber, Gabriel Puebla-Hellmann,
  Ute Drechsler, Tobias von Arx, Koushik Venkatesan, Lukas Novotny, and Emanuel
  L{\"{o}}rtscher.
\newblock {Fabrication of NEMS Actuated Plasmonic Antenna Platform for the
  Study of Optical Forces and Field Enhancements in Hotspots}.
\newblock In {\em Asia Communications and Photonics Conference 2016}. Optical
  Society of America, 2016.

\bibitem{Roxworthy2016}
Brian~J. Roxworthy and Vladimir~A. Aksyuk.
\newblock {Nanomechanical Motion Transduction with a Scalable Localized Gap
  Plasmon Architecture}.
\newblock {\em Nature Communications}, 7:13746, 2016.

\bibitem{Wilson2009}
Dalziel~J. Wilson, Cindy~A. Regal, Scott~B. Papp, and H.~J. Kimble.
\newblock {Cavity Optomechanics with Stoichiometric SiN Films}.
\newblock {\em Physical Review Letters}, 103(20):207204, 2009.

\bibitem{Yamada2013}
Shoko Yamada, Silvan Schmid, Tom Larsen, Ole Hansen, and Anja Boisen.
\newblock {Photothermal Infrared Spectroscopy of Airborne Samples with
  Mechanical String Resonators}.
\newblock {\em Analytical Chemistry}, 85(21):10531--10535, 2013.

\bibitem{Schmid2014a}
Silvan Schmid, Tolga Bagci, Emil Zeuthen, Jacob~M. Taylor, Patrick~K. Herring,
  Maja~C. Cassidy, Charles~M. Marcus, Luis {Guillermo Villanueva}, Bartolo
  Amato, Anja Boisen, Yong~Cheol Shin, Jing Kong, Anders~S. S{\o}rensen, Koji
  Usami, and Eugene~S. Polzik.
\newblock {Single-Layer Graphene on Silicon Nitride Micromembrane Resonators}.
\newblock {\em Journal of Applied Physics}, 115(5), 2014.

\bibitem{Gavartin2012}
Emanuel Gavartin, Pierre Verlot, and Tobias~J. Kippenberg.
\newblock {A Hybrid On-Chip Optomechanical Transducer for Ultrasensitive Force
  Measurements}.
\newblock {\em Nature Nanotechnology}, 7(8):509--514, 2012.

\bibitem{Larsen2013}
Tom Larsen, Silvan Schmid, Luis~Guillermo Villanueva, and Anja Boisen.
\newblock {Photothermal Analysis of Individual Nanoparticulate Samples Using
  Micromechanical Resonators}.
\newblock {\em ACS Nano}, 7(7):6188--6193, July 2013.

\bibitem{Lifshitz2010}
Ron Lishitz and M.C. Cross.
\newblock {\em Nonlinear Dynamics of Nanomechanical Resonators}.
\newblock Wiley-VCH, 2010.

\bibitem{Loebich1972}
Otto Loebich.
\newblock {The Optical Properties of Gold}.
\newblock {\em Gold Bulletin}, 5(1):2--10, March 1972.

\bibitem{Brand2015}
Olivier Brand, Isabelle Dufour, Stephen Heinrich, and Fabien Josse.
\newblock {\em Resonant MEMS: Fundamentals, Implementation and Applications}.
\newblock Wiley-Blackwell, 2015.

\end{thebibliography}

\end{document}